\documentclass[nofootinbib,twocolumn,aps,pre,superscriptaddress,citeautoscript,floatfix]{revtex4-2}
\setcitestyle{super}
\usepackage{graphics}
\usepackage{graphicx}
\usepackage{mathrsfs}
\usepackage{amsmath}
\usepackage{bm}
\usepackage{color}
\usepackage{float}

\begin{document}

\title{Formation of Diblock Copolymer Nanoparticles: Theoretical Aspects}
\author{Yanyan Zhu}
\affiliation{Center of Soft Matter Physics and its Applications, Beihang University, Beijing 100191, China}
\affiliation{School of Physics, Beihang University, Beijing 100191, China}
\author{Bin Zheng}
\affiliation{Raymond and Beverly Sackler School of Physics and Astronomy, Tel Aviv University, Ramat Aviv 69978, Tel Aviv, Israel}
\author{David Andelman}
\affiliation{Raymond and Beverly Sackler School of Physics and Astronomy, Tel Aviv University, Ramat Aviv 69978, Tel Aviv, Israel}
\author{Xingkun Man}
\email{manxk@buaa.edu.cn}
\affiliation{Center of Soft Matter Physics and its Applications, Beihang University, Beijing 100191, China}
\affiliation{School of Physics, Beihang University, Beijing 100191, China}

\begin{abstract}
We explore the shape and internal structure of diblock copolymer (di-BCP) nanoparticles (NPs) by using the Ginzburg-Landau free-energy expansion. The self-assembly of di-BCP lamellae confined in emulsion droplets can form either ellipsoidal or onion-like NPs. The corresponding inner structure is a lamellar phase that is either perpendicular to the long axis of the ellipsoids (L$_\perp$) or forms a multi-layer concentric shell (C$_\parallel$), respectively. We focus on the effects of the interaction parameter between the A/B monomers $\tau$, and the polymer/solvent $\chi$, as well as the NP size on the nanoparticle shape and internal morphology. The aspect ratio ($l_{\rm AR}$) defined as the length ratio between the long and short axes is used to characterize the overall NP shape. Our results show that for the solvent that is neutral towards the two blocks, as $\tau$ increases, the $l_{\rm AR}$ of the NP first increases and then decreases, indicating that the NP becomes more elongated and then changes to a spherical NP. Likewise, decreasing $\chi$ or increasing the NP size can result in a more elongated NP. However, when the solvent has a preference towards the A or B blocks, the NP shape changes from striped ellipsoid to onion-like sphere by increasing the A/B preference parameter strength. The critical condition of the transition from an L$_\perp$ to C$_\parallel$ phase has been identified. Our results are in good agreement with previous experiments, and some of our predictions could be tested in future experiments.
\end{abstract}

\maketitle

\section{Introduction}

\begin{figure*}[h!t]
{\includegraphics[width=0.7\textwidth,draft=false]{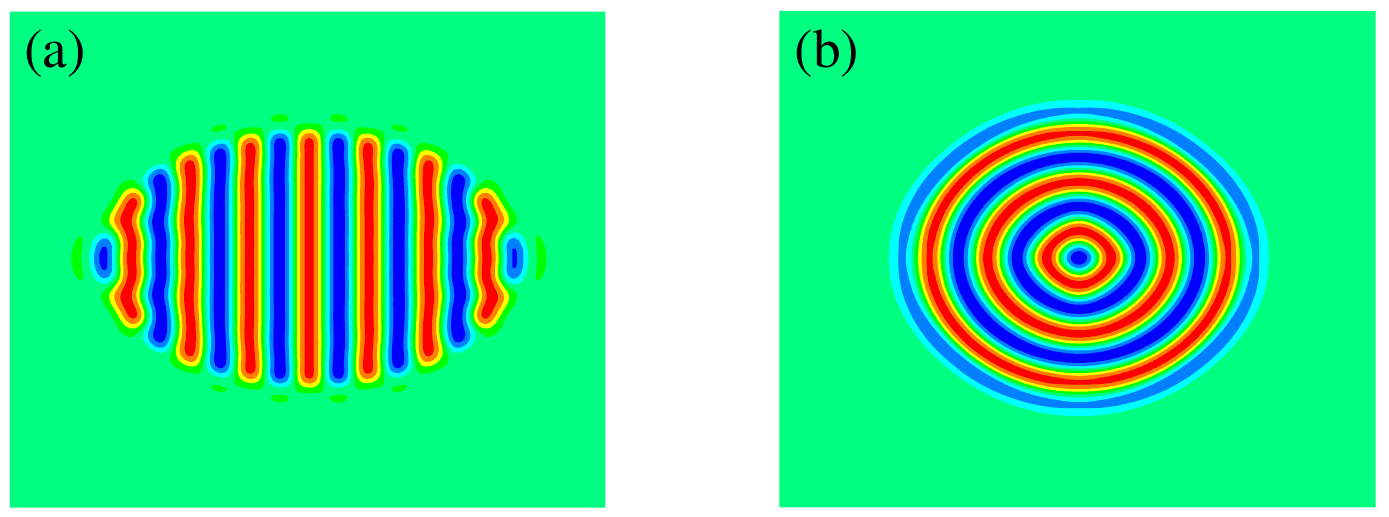}}
\caption{
\textsf{Two typical microphases of A/B di-block copolymer nanoparticles (BCP-NPs) embedded in a poor solvent solution. (a)~Striped ellipsoidal NP. (b)~Onion-like spherical NP. A-rich lamellae are colored red and B-rich are blue.}}
\label{figtf}
\end{figure*}

Polymeric nanoparticles (NPs) with well-controlled internal structure have attracted a large amount of interest due to their multiple applications~\cite{Shi13,Ku18,Shin20}. The self-assembly of block copolymers (BCP) into droplets has been proven to be a useful method to produce NPs with well-defined shapes and internal morphologies~\cite{Jeon08,Wyman11,Kim15}. Such NPs containing BCP (BCP-NPs) are used in a wide range of applications such as drug delivery, smart coatings, catalysis and optical lenses ~\cite{Alberto05,Deng13,Ku13,Ku14}. From a fundamental viewpoint, the formation of BCP-NPs is related to the minimization of the interfacial energy of soft particles during the self-assembly process. Consequently, diverse particle shapes are formed under soft confinement, including onion-like spherical NPs~\cite{Saito07,Tanaka09}, striped ellipsoidal NPs~\cite{Jang13,Lee20}, conical shaped NPs~\cite{Deng14,Kim19}, oblate NPs composed of cylindrical phases~\cite{Yang16,Lee19} and more. Here, the soft confinement refers to that the interface between NP and surrounding solvent is soft and can be easily deformed. We illustrate in Figure~\ref{figtf} two of the most common shapes of BCP-NPs that will be used throughout our paper: striped ellipsoidal shape (a); and, an onion-like spherical shape (b).

One of the current experimental challenges is to fabricate ellipsoidal BCP-NPs with precisely controlled shapes characterized by the particle aspect-ratio ($l_{\rm AR}$) and internal morphology. As our present study is restricted in two dimensions (2D), the aspect ratio, $l_{\rm AR}$, is defined as the ratio between the long and short axes of the ellipsoidal particle. Significant efforts have been devoted in the past to address this fabrication challenge. Among experimental attempts to control the particle shape, we mention the addition of surfactants, inclusion of other solid-like NPs, and varying the solvent evaporation rate~\cite{Deng132,Ku15,Lee17,Shin17}.

Kim and co-workers conducted a series of experiments on BCP-NPs. They systematically investigated the fabrication of such NPs with various shapes and inner structures. They found that the NP size, BCP molecular weight~\cite{Shin18} as well as the Flory-Huggins interaction parameter~\cite{Shin19} are three key parameters that control the shape ($l_{\rm AR}$) of ellipsoidal BCP-NPs. Moreover, the addition of surfactants, including length-controlled nanorod surfactants~\cite{Ku182} or photoresponsive surfactants~\cite{Lee192}, can induce the transition of striped ellipsoidal NPs to onion-like spherical NPs, due to the variation of the interfacial energy between BCP-NPs and solvent. On the theoretical side, phenomenological free energies of BCP-NPs have been proposed~\cite{Jang13,Klinger14,Ku19} to explain these experimental findings. It was shown that the NP shapes are determined by the competition between (i) the interfacial energy between the A/B di-BCP, (ii) the entropic penalty due to chain stretching and bending, and (iii) the surface energy of BCP-NPs with its surrounding medium (poor solvent).

In a separate study, Avalos et al.~\cite{Avalos18} presented experimental results of annealing BCP-NPs and developed a theoretical model based on the Cahn-Hilliard model to describe the morphological evolution of striped ellipsoids into onion-like spheres (see Figure~\ref{figtf}). Their numerical results showed that the annealing process changes the temperature dependence of the Flory-Huggins parameter and the interface width, and eventually results in a morphology change. Chi et al.~\cite{Chi11} studied the BCP self-assembly in soft confinement in poor solvent conditions using the simulated annealing method. They predicted  different shapes and internal structures of BCP-NPs as function of the polymer-solvent interaction strength and BCP monomer concentration in solution.

The formation of BCP-NPs is easily obtained in experiments, but at present, there is a lack of rigorous theoretical description. Hence, a systematic theoretical study of the formation mechanism of BCP-NPs, and, in particular, the influence of these key experimental parameters (mentioned above) on the NP shape, are still necessary. The theoretical study will not only enhance our fundamental understanding of how these parameters determine the final shape of BCP-NPs, but may also aid in the development of applications.

To address further this challenge, we employ a Ginzburg-Landau free energy and study the self-assembly of BCP-NPs embedded in a poor solvent. Particularly, we focus on the effect of the A/B block interaction parameter, the NP size, and the symmetric as well as asymmetric interactions between the polymer blocks and the solvent on the final shape and inner structure of BCP-NPs. We show that most of the experimental results of lamellar-forming BCP-NPs can be captured by our model. Furthermore, we present several predictions on how experimental controllable parameters affect the final shape of the BCP-NPs.

Hereafter, our model based on the Ginzburg-Landau free energy is introduced in section II. In section III, the calculated phase-diagram separating NPs of distinct shapes is presented in terms of the interactions between different system components and the NP size. Finally, section IV contains a comparison between existing experiments and other models, followed, in section V, by the some conclusions and future prospects.

\section{Model}

We study the equilibrium shape of di-block copolymer NPs (BCP-NPs) embedded in a solution by using the Ginzburg-Landau free energy~\cite{Netz97,Cohen14}. We model a mixture of three components: the A and B components of di-BCPs and the solvent. The incompressibility condition is $\phi_{\mathrm{A}}~+~\phi_{\mathrm{B}}~+~\phi_{\mathrm{S}}=1$, where $\phi_{\mathrm{A}}$, $\phi_{\mathrm{B}}$, and $\phi_{\mathrm{S}}$ are the volume fractions of the A, B blocks and solvent, respectively.

Two order parameters are introduced for convenience,
\begin{equation}
\begin{aligned}
&\rho=\phi_{\mathrm{A}}+\phi_{\mathrm{B}},  \\
&\phi=\phi_{\mathrm{A}}-\phi_{\mathrm{B}}
\end{aligned}
\end{equation}
where $\rho$ is the total BCP volume fraction, and $\phi$ is the concentration difference between the A and B blocks.

In the grand-canonical ensemble, the Ginzburg-Landau free energy of the A/B BCP solution is written as~\cite{Cohen14}:
\begin{equation}
\begin{aligned}
\frac{g(\phi, \rho)}{k_{\mathrm{B}} T}=&-\frac{\tau}{2} \phi^{2}+\frac{\chi}{2} \rho(1-\rho)+ \nu_{\phi \rho} \phi(1-\rho)-\mu_{\phi} \phi-\mu_{\rho} \rho \\
&+\frac{\rho+\phi}{2} \ln (\rho+\phi)+\frac{\rho-\phi}{2} \ln (\rho-\phi)+(1-\rho) \ln (1-\rho) \\
&+\frac{H}{2}\left[\left(\nabla^{2}+q_{0}^{2}\right) \phi\right]^{2}+K(\nabla \rho)^{2}
\end{aligned}
\end{equation}
In the above free energy, the first three terms are the interaction terms, where $\tau$ is the interaction parameter between the A and B monomers, $\chi$ is the interaction parameter between solvent and polymer, and $\nu_{\phi \rho}$ is the parameter denoting the preference interaction of solvent towards either the A or B monomers, and $\mu_{\phi}$ and $\mu_{\rho}$ are the corresponding chemical potentials. The next three terms are the ideal entropy of mixing, while the $H$-term induces a modulated phase in the A/B relative concentration $\phi$. The $H$ prefactor is the modulation parameter and $q_{0}=1/\sqrt{2}$ (in dimensionless units) is the most dominant wavenumber~\cite{Fredrickson87}. The last item, parameterized by the $K$ prefactor, is used to characterize the stiffness of the interface between the polymer and the solvent~\cite{Safran03}.

In order to analysis the formation mechanism of BCP ellipsoidal NPs, we focus on two separate energy terms of the full Ginzburg-Landau free energy, eq~(2). The first one is the polymer free energy $F_{\rm P}$, which includes both enthalpy and entropy of the polymer chains,
\begin{equation}
\begin{aligned}
F_{\rm P}=-\frac{\tau}{2} \phi^{2}+\frac{\rho+\phi}{2} \ln (\rho+\phi)+\frac{\rho-\phi}{2} \ln (\rho-\phi)
\end{aligned}
\end{equation}
The second term is the polymer-solvent interfacial energy $E_{\rm PS}$,
\begin{equation}
\begin{aligned}
E_{\rm PS}=\frac{\chi}{2} \rho(1-\rho)+ \nu_{\phi \rho} \phi(1-\rho)
\end{aligned}
\end{equation}
We will show that the BCP-NP equilibrium shape is mainly determined by the competition between these two terms of the full free energy, eq~(2).

We use the conjugate gradient (CG) method to minimize the Ginzburg-Landau free energy, eq~(2), in order to obtain the equilibrium density distribution of each component. The numerical calculations are performed in a two-dimensional (2D) box ($L_{x}\times L_{y}$) with a spatial grid $\Delta x=\Delta y=0.05$. In our simulations, all lengths are expressed in units of the bulk natural periodicity of the lamellar phase, $L_0 \equiv 2\pi/q_0$.

\section{Results}

From experiments, it is known that the lamellar phase of BCP solution can form either ellipsoidal or onion-like NPs (Figure~\ref{figtf}). The lamellar order is usually perpendicular to the long axis for ellipsoidal NPs, while for onion-like NPs they form concentric shell structures. The BCP-NP final shape is mainly determined by the A/B monomer interaction parameter $\tau$, the polymer/solvent interaction parameter $\chi$, the preference of the solvent towards the two components $\nu_{\phi \rho}$, and the size of the NPs. Therefore, we mainly focus on the effects of these experimental controllable parameters on the shape and inner structure of the BCP-NPs.

\subsection{Ellipsoidal Particles (L$_\perp$)}

When the solvent preference towards the A and B components is neutral, $\nu_{\phi \rho}=0$, ellipsoidal NPs are usually formed. In this subsection, we present a systematical study of the effect of the parameters $\tau$, $\chi$, and the NP size on the final shape of the BCP-NPs, where for ellipsoidal NPs, the aspect ratio $l_{\rm AR}$ serves as the shape parameter and is defined as the length ratio between the major and minor axes.

\subsubsection{The effect of $\tau$ and $\chi$ on the NP shape}

\begin{figure*}[h!t]
{\includegraphics[width=1.0\textwidth,draft=false]{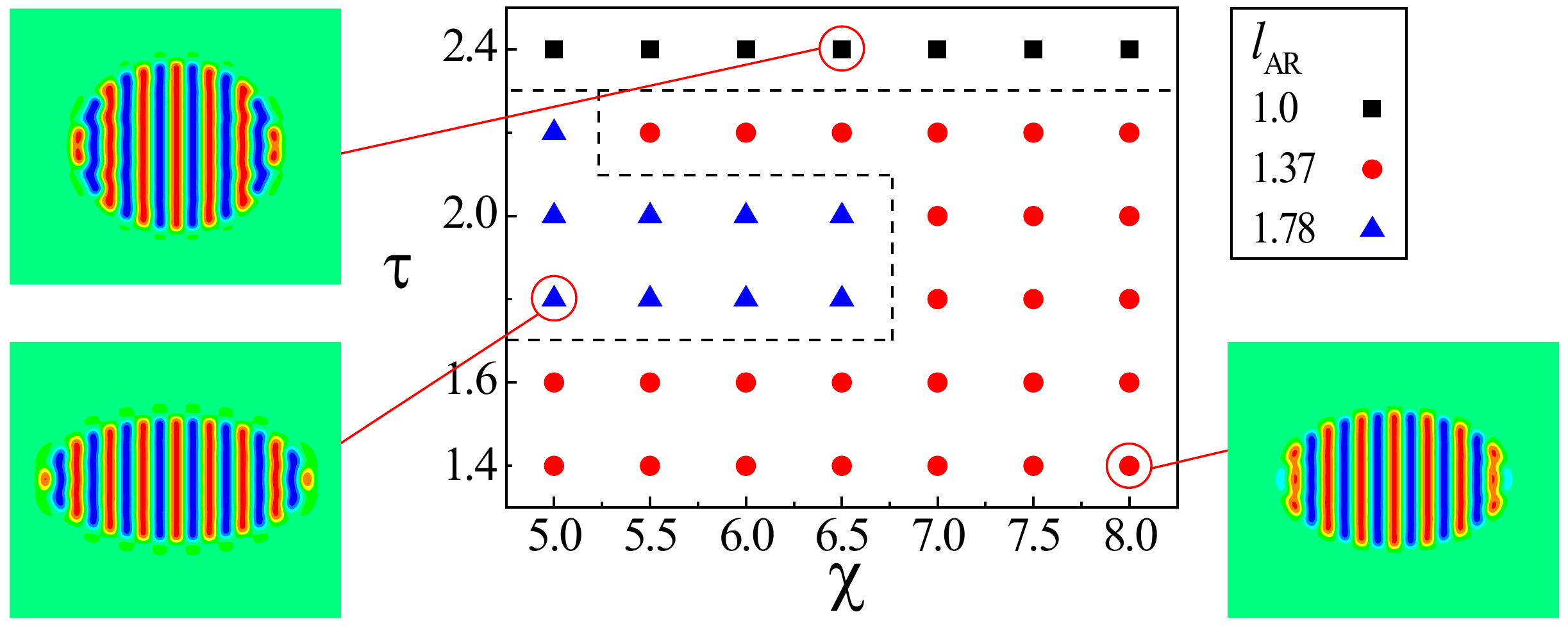}}
\caption{
\textsf{The phase diagram plotted in the ($\chi$, $\tau$) parameter plane for a BCP-NP embedded in a poor-solvent solution. The three distinct shapes denote the different values of the aspect ratio,  $l_{\rm AR}$, that minimize the free energy. The black square corresponds to an NP with $l_{\rm AR}=1$, red circle corresponds to $l_{\rm AR}=1.37$, and blue triangle corresponds to $l_{\rm AR}=1.78$ (see legend). The box size is $10\times10$ in units of $L_0=2\pi/q_0$. Other parameter values are: $\nu_{\phi\rho}=0$, $H=6$, and $K=0.5$.}}
\label{figpd}
\end{figure*}

We first present the effect of the interaction parameter $\tau$, between the A and B components, and the interaction parameter $\chi$ between polymer and solvent, on the equilibrated $l_{\rm AR}$ of the BCP-NPs. Figure~\ref{figpd} shows the phase diagram in terms of distinct values of the $l_{\rm AR}$ in the ($\chi$, $\tau$) plane. For a given set of ($\chi$, $\tau$), we calculate the free energy as a function of $l_{\rm AR}$ by varying $l_{\rm AR}$ in the interval $[0.69, 2.25]$. The equilibrium $l_{\rm AR}$ value is always associated with the free energy at its minimum.

As can be seen in the phase diagram of Figure~\ref{figpd}, for $\chi \ge 7.0$, $l_{\rm AR}$ nearly does not change as long as $\tau<2.4$. When $\tau \ge 2.4$, the NPs tends towards a spherical (circular in 2D) shape. This indicates that in the large $\chi$ region, large enough values of $\tau$ can decrease $l_{\rm AR}$. For $\chi<7.0$, the dependence of $l_{\rm AR}$ on $\tau$ becomes non-monotonous. In this $\chi$ region, for fixed $\chi$ and by increasing $\tau$, $l_{\rm AR}$ first increases and then decreases. The phase diagram also shows that when $1.8 \le \tau \le 2.2$, the NP has a reduced $l_{\rm AR}$ value when $\chi$ increases. On the other hand, when $\tau$ lies outside the above-mentioned range, $\chi$ has no effect on $l_{\rm AR}$. For smaller value of $\tau<1.8$, the NP has an ellipsoidal shape with a fixed $l_{\rm AR}$, and for large values of $\tau>2.2$, the NP has a spherical shape for all calculated $\chi$ values.

\begin{figure*}[h!t]
{\includegraphics[width=0.95\textwidth,draft=false]{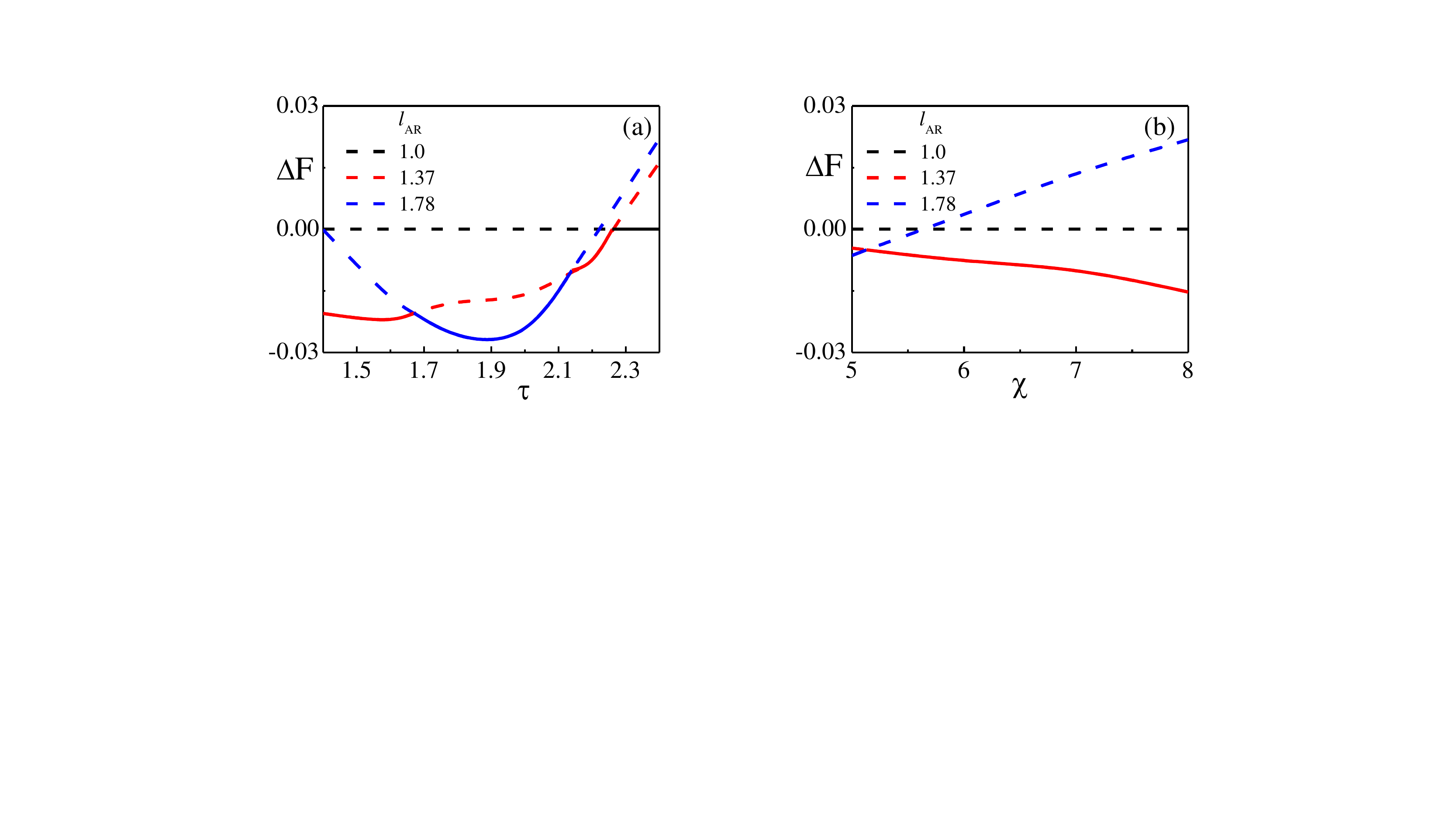}}
\caption{
\textsf{(a)~Dependence of the free energy difference, $\Delta F\equiv F(l_{\rm AR})-F(l_{\rm AR}=1)$ on the A/B monomer interaction parameter $\tau$, for $l_{\rm AR}=1$ (black line), $l_{\rm AR}=1.37$ (red line) and $l_{\rm AR}=1.78$ (blue line), as $\chi=5.5$. The solid line indicates which of the three lines is the stable one. (b)~Dependence of the free energy difference, $\Delta F$, on the polymer-solvent interaction parameter, $\chi$, for $l_{\rm AR}=1$ (black line), $l_{\rm AR}=1.37$ (red line) and $l_{\rm AR}=1.78$ (blue line), as $\tau=2.2$. The box size is $10\times 10$ in units of $L_0=2\pi/q_0$. Other parameter values are: $\nu_{\phi\rho}=0$, $H=6$, and $K=0.5$.}}
\label{figmf}
\end{figure*}

Two typical transitions mentioned above can also be seen in Figure~\ref{figmf}. In Figure~\ref{figmf}a, we present the three $l_{\rm AR}$ branches of $\Delta F$, the minimal one as $\tau$ varies can be identified, and $\Delta F$ is defined as the difference between the actual free energy and the reference free energy of $l_{\rm AR}=1$, for three values of $l_{\rm AR}$. Figure~\ref{figmf}b shows that the $l_{\rm AR}$ determined by the free energy minimization changes from $l_{\rm AR}=1.78$ to $l_{\rm AR}=1.37$ as $\chi$ increases from $\chi =5.0$ to $\chi=8.0$.

\begin{figure*}[h!t]
{\includegraphics[width=1.0\textwidth,draft=false]{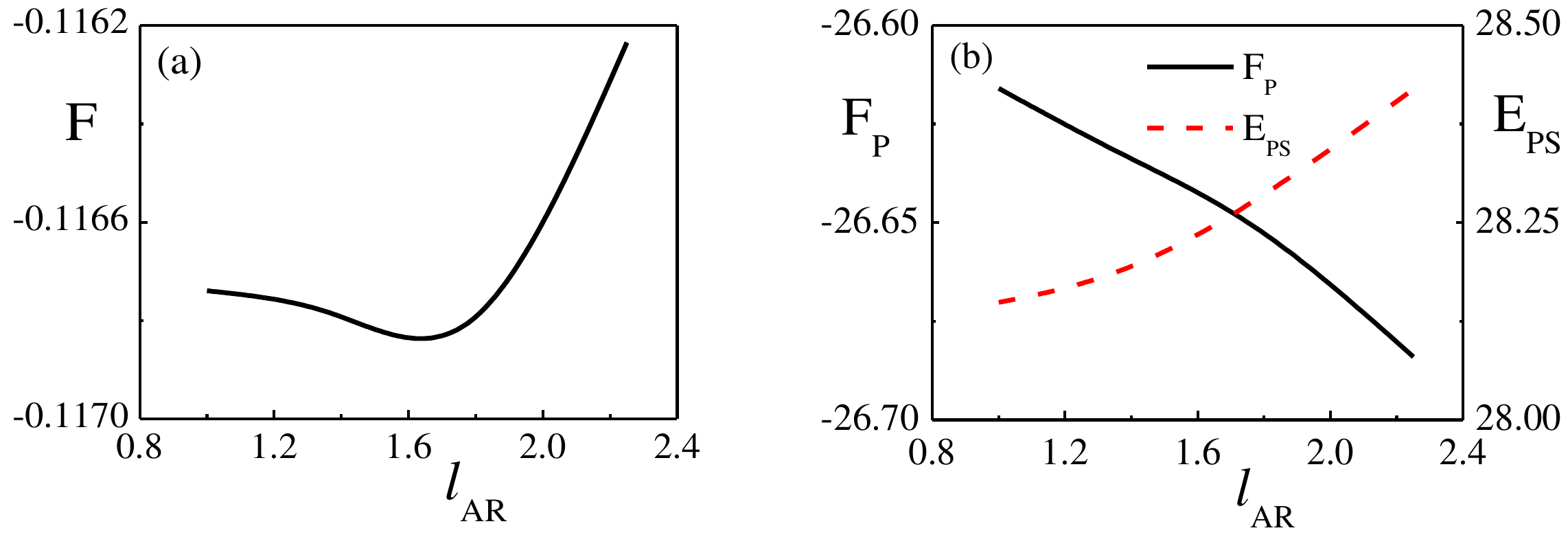}}
\caption{
\textsf{(a)~Dependence of the free energy $F$ on the aspect ratio, $l_{\rm AR}$. (b)~Separate plots of the polymer energy $F_{\rm P}$ (solid black line) and interfacial energy $E_{\rm PS}$ (dashed red line) on the aspect ratio $l_{\rm AR}$. The $F_{\rm P}$ and $E_{\rm PS}$ energy terms are plotted as solid black and dashed red lines, respectively. For all cases, $\tau=2.2$ and $\chi=5.0$. The box size is $10\times 10$ in units of $L_0=2\pi/q_0$. Other parameter values are: $\nu_{\phi\rho}=0$, $H=6$, and $K=0.5$.}}
\label{figlar}
\end{figure*}

Figure~\ref{figlar}a shows that when $\tau=2.2$ and $\chi=5.0$, the free energy has a minimum at $l_{\rm AR}=1.78$, indicating the formation of ellipsoidal NPs.  To understand why the NPs have an equilibrium $l_{\rm AR}$ for a given ($\chi$, $\tau$), we calculate the two separate contributions to the total free energy, $F_{\rm P}$ and $E_{\rm PS}$, as function of $l_{\rm AR}$ and show it on Figure~\ref{figlar}b. Here, $F_{\rm P}$ is the polymer free energy, including the enthalpy and entropy, and $E_{\rm PS}$ is the polymer/solvent interfacial energy. Our results indicate that for fixed values of the ($\chi$, $\tau$) parameters, as $l_{\rm AR}$ increases, $F_{\rm P}$ decreases but $E_{\rm PS}$ increases. The decreased in $F_{\rm P}$ is due to the orientation of the lamellae perpendicular to the long axis. The BCPs feel the confinement only along the long axis direction, and not along the short axis.

One of the conclusions is that the confinement becomes weaker when the long axis of the ellipsoid becomes longer. Namely, larger $l_{\rm AR}$ results in less confinement to the BCP lamellae. However, when $l_{\rm AR}$ increases, the polymer/solvent interfacial energy increases, resulting in an increase of $E_{\rm PS}$. In other words, $F_{\rm P}$ tends to elongate the NP, while $E_{\rm PS}$ prefers to make it more spherical. The balance between these two terms determines the final shape of the NP.
Figure~\ref{figlar}b shows that $F_{\rm P}$ is always negative, therefore, the effect of $F_{\rm P}$ on the NP shape is stronger when its value is smaller. On the other hand, as $E_{\rm PS}$ is always positive, its constraining effect on the final shape is stronger when its value is larger.

After understanding why the system has a preferred aspect ratio, $l_{\rm AR}$, for a given set of $\tau$ and $\chi$, we turn to the non-monotonic effect of $\tau$ on the aspect ratio, $l_{\rm AR}$. The phase diagram of Figure~\ref{figpd} shows that as $\tau$ increases from $\tau=1.4$ to $\tau=2.4$ (and for fixed $\chi=5.5$), $l_{\rm AR}$ first increases from $l_{\rm AR}=1.37$ to $l_{\rm AR}=1.78$, then it goes back to $l_{\rm AR}=1.37$, and finally its value drops to $l_{\rm AR}=1.0$.

\begin{figure*}[h!t]
{\includegraphics[width=1.0\textwidth,draft=false]{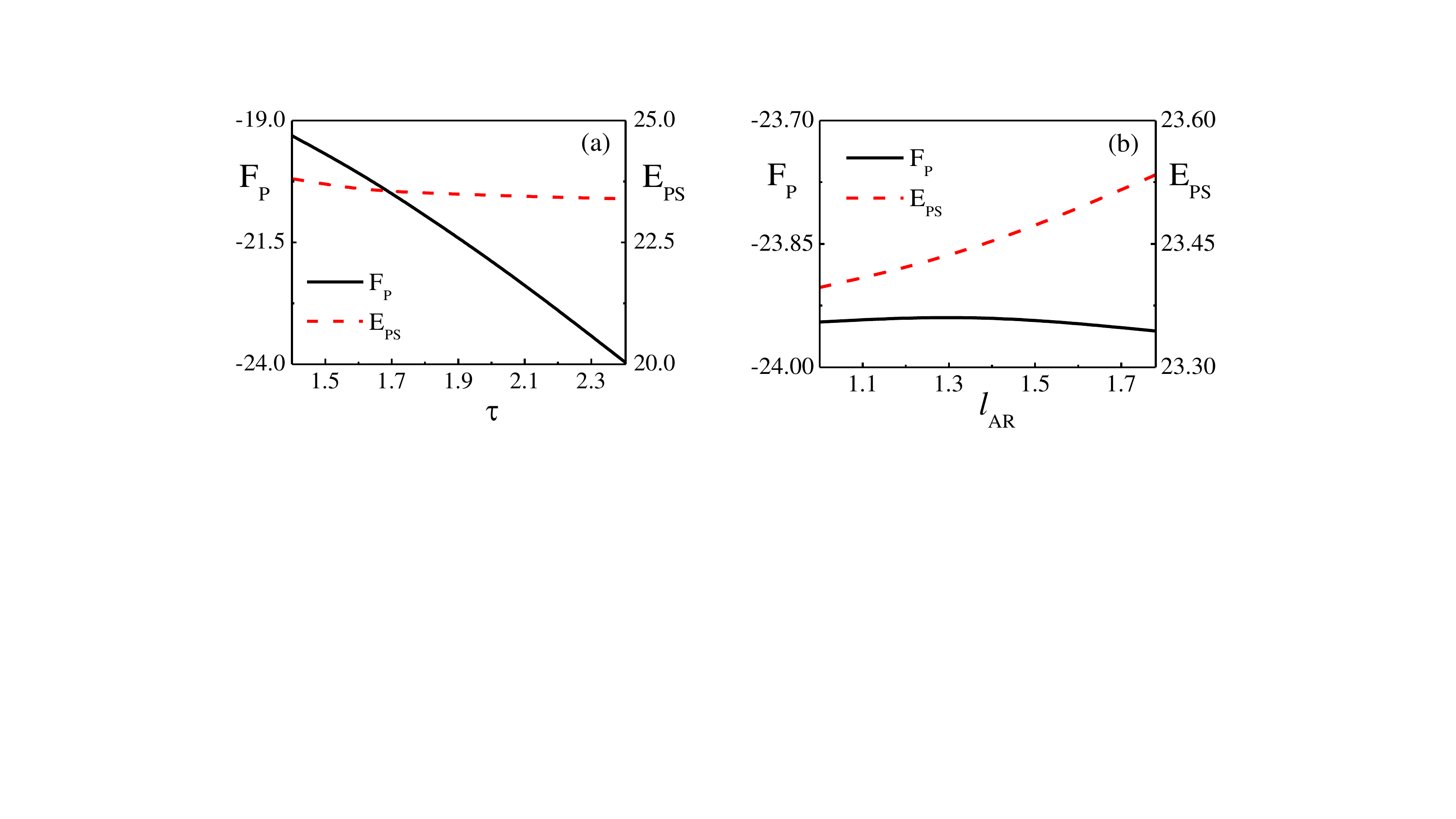}}
\caption{
\textsf{(a)~The dependence of $F_{\rm P}$ (solid black line) and $E_{\rm PS}$ (dashed red  line) on $\tau$ for $l_{\rm AR}=1$. (b)~The dependence of $F_{\rm P}$ (solid black line) and $E_{\rm PS}$ (dashed red line) on $l_{\rm AR}$ for $\tau=2.4$. In all cases, $\chi=5.5$ and the box size is $10\times 10$ in units of $L_0=2\pi/q_0$. Other parameter values are: $\nu_{\phi\rho}=0$, $H=6$, and $K=0.5$.
}}
\label{figtau}
\end{figure*}

We analysis such non-monotonic behavior by exploring the dependence of $E_{\rm PS}$ and $F_{\rm P}$ on $\tau$ in Figure~\ref{figtau}a, where it is shown that both $E_{\rm PS}$ and $F_{\rm P}$ decrease as $\tau$ increases. This indicates that increasing $\tau$ enhances the elongation effect of $F_{\rm P}$ on the NP shape, but it also weakens the constraining effect of $E_{\rm PS}$ that causes the NP to be more spherical. Therefore, as $\tau$ increases, $F_{\rm P}$ wins over the second term and overall the NP  becomes more elongated.

When $\tau$ is large enough (for example $\tau=2.4$), the change in $l_{\rm AR}$ is caused by a different mechanism. Figure~\ref{figtau}b shows the dependence of $F_{\rm P}$ and $E_{\rm PS}$ on $l_{\rm AR}$ for $\tau=2.4$. In this case, $F_{\rm P}$ is nearly a constant when $l_{\rm AR}$ varies, and can be explained in the following way. For large enough $\tau$, the BCPs within the NP undergo strong segregation, and the polymer free-energy depends mainly on the enthalpy term and not on the entropy one. Therefore, $F_{\rm P}$ has only a weak dependence on the NP shape. However, $E_{\rm PS}$ has a relatively large change when $l_{\rm AR}$ increases from $l_{\rm AR}=1$ to $l_{\rm AR}=1.78$. This indicates that for large $\tau$ (strong segregation), the NP shape is mainly determined by the interface energy $E_{\rm PS}$, causing the NP to be more spherical.

The effect of $\chi$ on $l_{\rm AR}$ manifests itself in two different ways depending on the specific value of $\tau$. For intermediate values, $1.8\le \tau \le 2.2$, increasing $\chi$ decreases $l_{\rm AR}$. However, for $\tau$ outside this intermediate range, the effect of $\chi$ on $l_{\rm AR}$ is weak, resulting in almost a constant value of $l_{\rm AR}$ even when the $\chi$ value varies. As is discussed above, when $\tau$ is large enough, the NP shape is mainly determined by the interfacial energy. Therefore, for the large value of $\tau=2.4$, the NP is spherical in a wide range of $\chi$ values.

\begin{figure*}[h!t]
{\includegraphics[width=0.95\textwidth,draft=false]{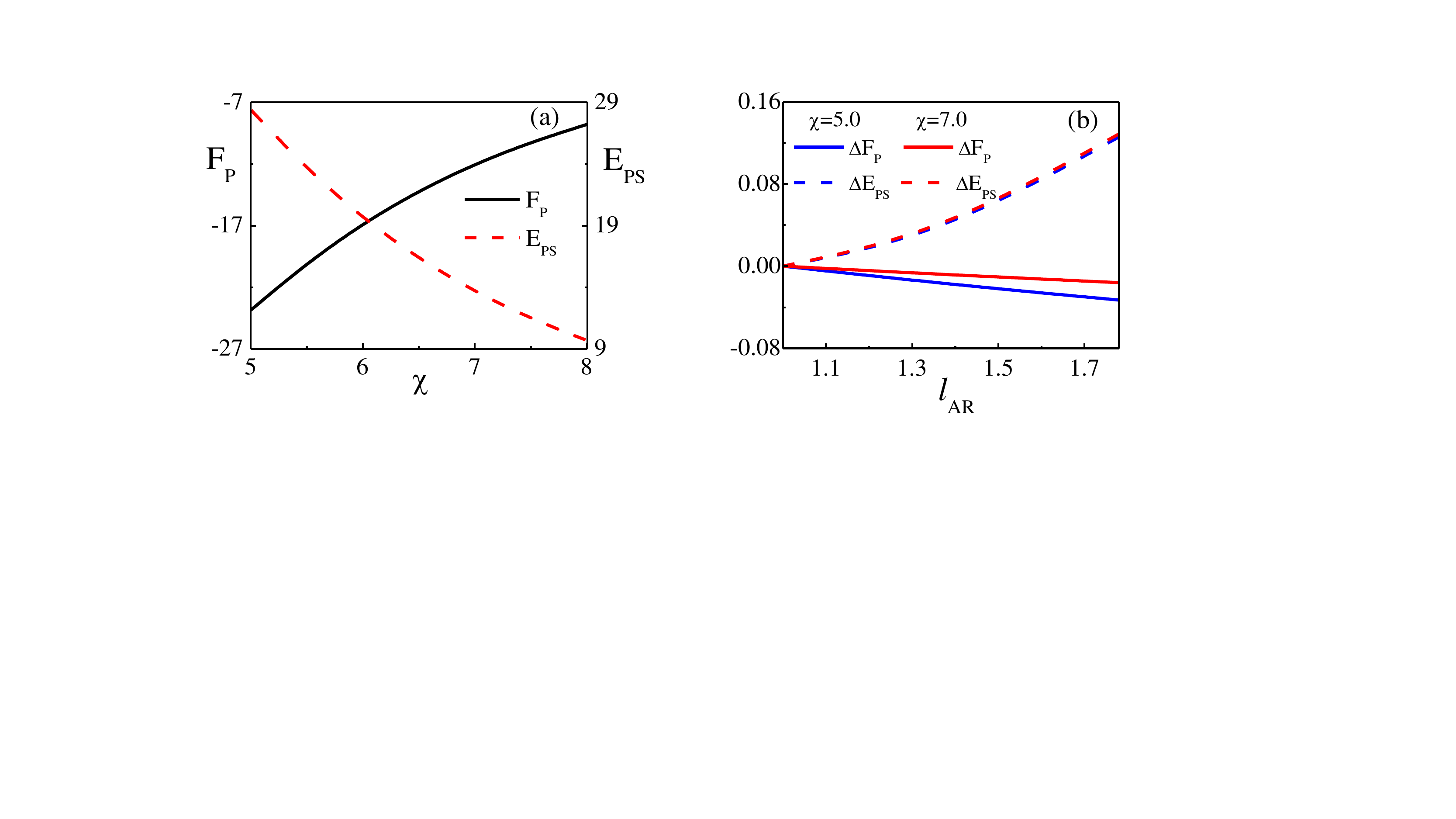}}
\caption{
\textsf{(a)~$F_{\rm P}$ (solid black line) and $E_{\rm PS}$ (dashed red line) as function of the polymer-solvent interaction parameter, $\chi$, as $\tau=1.6$. (b)~Dependence of $\Delta{F_{\rm P}}$ (solid line) and $\Delta{E_{\rm PS}}$ (dashed line) on $l_{\rm AR}$ for $\chi=5.0$ (blue line) and $\chi=7.0$ (red line) as $\tau=2.2$. For all cases,  the box size is $10\times10$ in units of $L_0=2\pi/q_0$, and $\nu_{\phi\rho}=0$, $H=6$, and $K=0.5$.
}}
\label{figchi}
\end{figure*}

For smaller $\tau$ values, the dependency of $E_{\rm PS}$ and $F_{\rm P}$ on $\chi$ is shown in Figure~\ref{figchi}a. It elucidates that increasing $\chi$ increases $F_{\rm P}$ and decreases $E_{\rm PS}$. Hence, by increasing $F_{\rm P}$, the NP becomes more spherical, while decreasing $E_{\rm PS}$ makes the NP more elongated. By varying the $\chi$ value, the change in the two energy terms is almost the same, {\it i.e.}, $E_{\rm PS}$ decreases, while $F_{\rm P}$ increases by the same order of magnitude, resulting in a balance between these two energies. This is the reason why the minimum free-energy gives a fixed $l_{\rm AR}$ for the NP shape.

For intermediate values of $\tau$ and for $\chi=5.0$ and $7.0$, the dependence of $\Delta F_{\rm P}$ and $\Delta E_{\rm PS}$ on $l_{\rm AR}$ are shown in Figure~\ref{figchi}b, where $\Delta{F_{\rm P}}=F_{\rm P}(l_{\rm AR})-F_{\rm P}(l_{\rm AR}=1)$, is the difference between $F_{\rm P}(l_{\rm AR})$ and the reference state of $F_{\rm P}(l_{\rm AR}=1)$, and quite similarly $\Delta{E_{\rm PS}}$ is defined. Figure~\ref{figchi}b shows that $\Delta E_{\rm PS}$ for $\chi=5.0$ is nearly the same as for $\chi=7.0$. However, for $\chi=5.0$, $\Delta F_{\rm P}$ is smaller than its value for $\chi=7.0$. This indicates that the effect of the interfacial energy $E_{\rm PS}$ on $l_{\rm AR}$ is almost the same when $\chi$ increases, but the effect of $F_{\rm P}$ is a little weaker for $\chi=7.0$ than $\chi=5.0$. This is the reason why the NP shape becomes less elongated when $\chi$ increases.

\subsubsection{The size effect on the NP shape}

\begin{figure*}[h!t]
{\includegraphics[width=0.86\textwidth,draft=false]{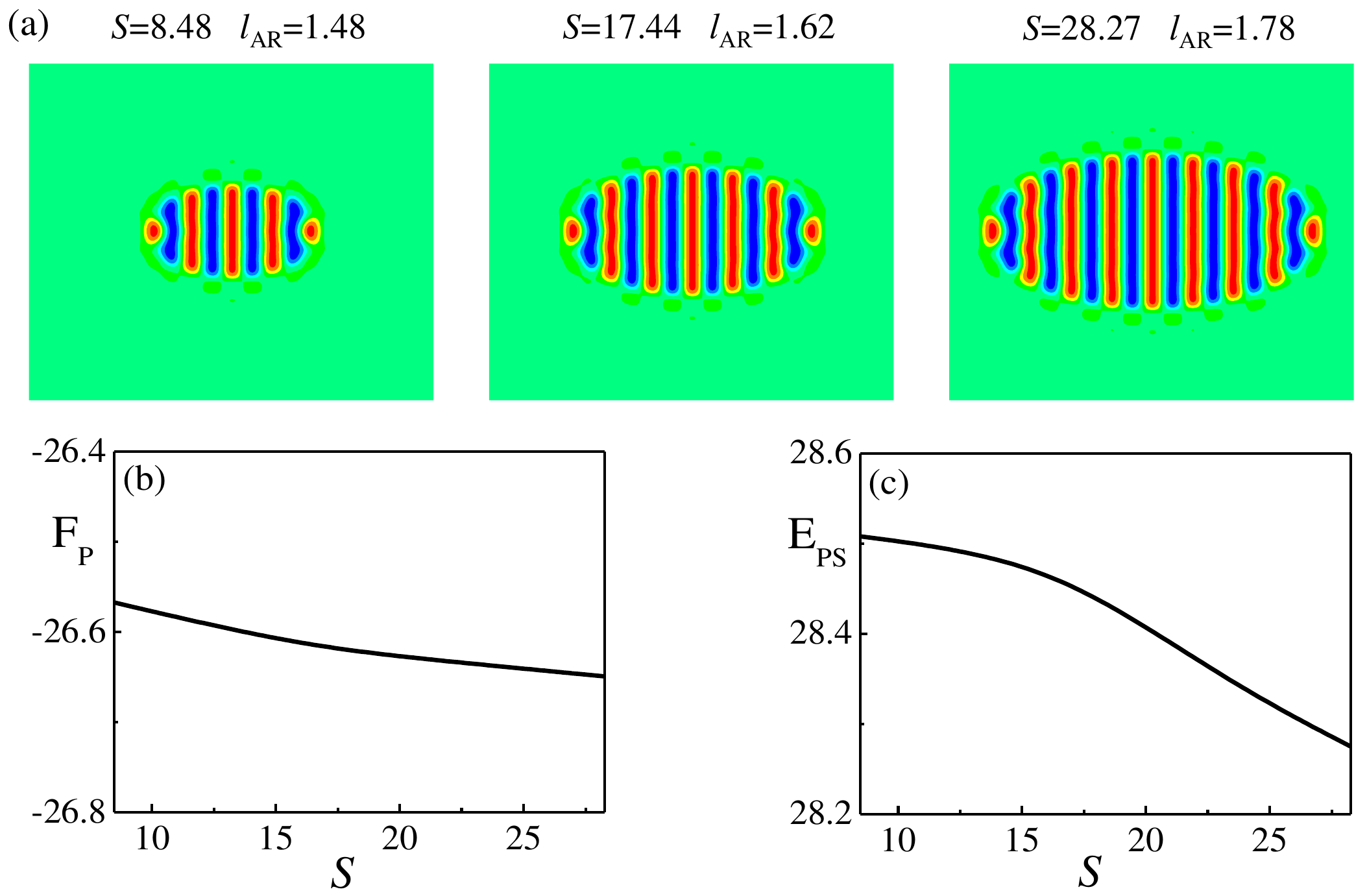}}
\caption{
\textsf{(a)~The equilibrated NP shapes for different areas, $S$. The corresponding area and aspect ratio from left to right are: $S=8.48$, $l_{\rm AR}=1.48$, $S=17.44$, $l_{\rm AR}=1.62$ and $S=28.27$, $l_{\rm AR}=1.78$. (b)~ and (c)~show the dependence of $F_{\rm P}$ and $E_{\rm PS}$ on the NP area $S$, respectively. For all cases, $\tau=2.2$ and $\chi=5.0$. The box size is $10\times10$ in units of $L_0=2\pi/q_0$, and $\nu_{\phi\rho}=0$, $H=6$, and $K=0.5$.
}}
\label{figsize}
\end{figure*}

In addition to varying $\tau$ and $\chi$, we also investigated the NP size effect on the final particle shape. Here, we use the area of the ellipse $S=\pi a b$, where $a$ and $b$ are the two principal axes, to characterize the NP size because our calculations are done in 2D. For a given area $S$, we calculated the free energy of BCP-NPs for various $l_{\rm AR}$. The $l_{\rm AR}$ that corresponds to the minimum free energy is taken as the equilibrium one. Figure~\ref{figsize}a shows three equilibrium shapes of BCP-NPs for $S=8.48$, $17.44$, and $28.27$ with three corresponding equilibrated $l_{\rm AR}$. The energy terms, $F_{\rm P}$ and $E_{\rm PS}$, are plotted in Figure~\ref{figsize}b and ~\ref{figsize}c, respectively, decrease their values as the NP area $S$ increases. This indicates that increasing the NP size enhances the effect of $F_{\rm P}$, but weaken the constrain effect of $E_{\rm PS}$ on the equilibrium shape. Therefore, $l_{\rm AR}$ increases as the size increases, as in clearly seen in Figure~\ref{figsize}a.

\subsection{Onion-like Spherical Particles}
So far we discussed the neutral solvent case, $\nu_{\phi\rho}=0$. The resulting NPs have an ellipsoidal shape (in 2D) with inner BCP lamellae that are perpendicular to the long axis (L$_\perp$). This phenomenon is also seen in experiments. Let us now consider the case when the solvent has a preference towards one of the two blocks, {\it i.e.,} $\nu_{\phi\rho}\neq 0$. Experiments~\cite{Lee192} have shown that when the solvent prefers, say, the B component, an onion-like structure (C$_\parallel$) with a B-shell on the NP perimeter is formed. Clearly, the opposite occurs if the solvent prefers the A component, and it will result in an onion-like structure (C$_\parallel$) with an A-shell on its perimeter, in agreement with our calculations. Next, we would like to investigate how the onion-like structure that appears when the preference $\nu_{\phi\rho}$ is large enough.

\begin{figure*}[h!t]
{\includegraphics[width=1.0\textwidth,draft=false]{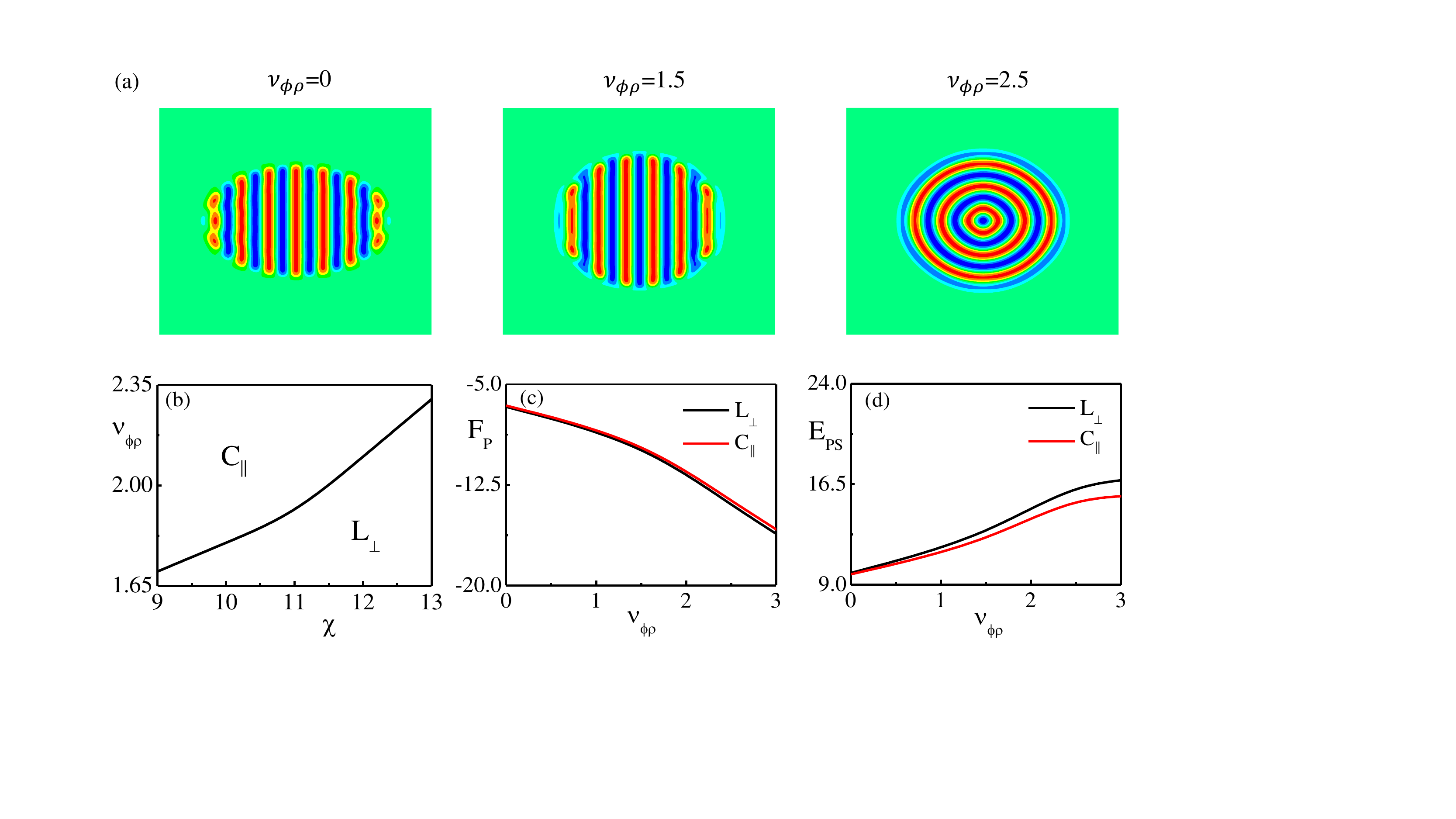}}
\caption{
\textsf{(a)~The equilibrium shape of an NP in increasing solvent preference. From left to right, the values of the $\nu_{\phi\rho}$ parameter are: $\nu_{\phi\rho}=0$, $1.5$, and $2.5$. (b)~The C$_\parallel$ to L$_\perp$ phase diagram in the ($\chi$, $\nu_{\phi\rho}$) plane for $\tau=1.4$. In (c)~and (d), $F_{\rm P}$ and $E_{\rm PS}$ are plotted as function of $\nu_{\phi\rho}$ for $\tau=1.4$ and $\chi=9.0$, respectively. The box size is $10 \times 10$ in units of $L_0=2\pi/q_0$. In addition,  $H=6$ and $K=3$.
}}
\label{fignu}
\end{figure*}

Besides the qualitative agreement with experiments, we find that $\nu_{\phi\rho}$ has a more interesting effect on the NP shape. Figure~\ref{fignu}a shows that by increasing the preference from $\nu_{\phi\rho}=0$ to $2.5$, the NP shape changes from an ellipsoidal to more spherical, and then to an onion-like sphere. At the same time, the inner structure changes from lamellae perpendicular to the long NP axis (L$_\perp$) to a structure with concentric shells (C$_\parallel$). For given values of $\tau$ and $\chi$, a critical value of $\nu_{\phi\rho}=\nu_{\phi\rho}^*$ is needed to induce the transition from L$_\perp$ to C$_\parallel$ structure. Figure~\ref{fignu}b shows that for $\tau=1.4$, the critical value $\nu_{\phi\rho}^*$ at the transition increases as $\chi$ increases. The transition line, shown in Figure~\ref{fignu}b, separates an upper C$_\parallel$ phase and an L$_\perp$ phase from below.

The change in the NP shape and inner structure as a function of $\nu_{\phi\rho}$ can also be observed by the dependency of $F_{\rm P}$ and $E_{\rm PS}$ on $\nu_{\phi\rho}$, as shown in Figures~\ref{fignu}c and \ref{fignu}d. Results show that $F_{\rm P}$ is always smaller for L$_\perp$ than for C$_\parallel$ structures. However, $E_{\rm PS}$ is always larger for the L$_\perp$ phase than for the C$_\parallel$. This means that the effect of $F_{\rm P}$ induces the formation of L$_\perp$ NPs, but $E_{\rm PS}$ leads to the formation of C$_\parallel$ with an onion-like inner structure. As $\nu_{\phi\rho}$ increases, the $E_{\rm PS}$ difference between L$_\perp$ and C$_\parallel$ becomes larger, and the NP shape changes from L$_\perp$ to C$_\parallel$. The spherical L$_\perp$ is the intermediate structure of this transition.

\section{Discussion}

\subsection{Comparison with Experiments}
Our study of the BCP-NP formation mechanism was inspired by a recent series of experimental studies~\cite{Shin18,Shin19}. Shin et al.~\cite{Shin18} have shown that the NP aspect-ratio $l_{\rm AR}$ can be controlled by tuning the NP size. They employed a membrane device made of Shirasu porous glass (SPG) to generate monodisperse NPs made from PS-b-PB block copolymers with different NP diameters. They found that the $l_{\rm AR}$ increases as the NP size increases, which is in excellent agreement with our predictions shown in Figure~\ref{figsize}. The dependence of $l_{\rm AR}$ on the NP size can be attributed to the fact that the increase in the NP size enhances the contribution of the bulk energy, and decreases the surface energy due to the reduced surface area/volume ratio. The bulk energy includes mainly the polymer free energy (this is the $F_{\rm P}$ in our calculations). This can be clearly seen in our work as increasing $l_{\rm AR}$ can decrease $F_{\rm P}$ (see Figure~\ref{figlar}b). For an L$_\perp$ NP, the lamellae only feel the confinement along the long axis direction. The larger $l_{\rm AR}$ values imply that the less confinement is applied to the BCP lamellar phase. In other words, lamellar BCPs prefer to form elongated NPs with inner lamellae perpendicular to the long axis (L$_\perp$).

Shin and co-workers~\cite{Shin19} applied various type of di-BCP to generate NPs each having a different Flory-Huggins parameter, $\tau$. Their study included PS$_{34\rm k}$-b-PB$_{25\rm k}$, PS$_{16\rm k}$-b-PDMS$_{17\rm k}$ and PS$_{10\rm k}$-b-P4VP$_{10\rm k}$ with $\tau$ getting values of  $\tau=0.04$, $0.21$ and $0.53$, respectively. They found that $l_{\rm AR}$ increases for the sequence of PS$_{34\rm k}$-b-PB$_{25\rm k}$, PS$_{16\rm k}$-b-PDMS$_{17\rm k}$, and PS$_{10\rm k}$-b-P4VP$_{10\rm k}$. Namely, increasing $\tau$ results in an increase of $l_{\rm AR}$. Our results not only confirms this finding, but also predicts that the NP becomes more spherical again for larger values of $\tau$. This prediction can be tested in experiments if the used di-BCPs will cover a larger range of equivalent $\tau$ values.

\subsection{Comparison with Other Models}

Our results concerning the dependence of NP shape on $\tau$ and on the NP size are in qualitative agreement with the theoretical analysis of Ku et al.~\cite{Ku19} that was based on a phenomenological coarse-grained free energy. In their work, the proposed free energy is composed of four terms: the interfacial energy between A/B blocks, the stretching energy of di-BCP chains, the bending energy of curved lamellae, as well as the surface energy at the NP/solvent interface. The first three terms are the bulk energy, while the last term is the polymer/solvent interfacial energy. Such a phenomenological free-energy shows that when $l_{\rm AR}$ increases, the A/B interfacial energy and the bending energy decrease, but the polymer/solvent surface energy increases. Moreover, when the NP size increases, the bulk energy increases, while the last surface energy decreases, resulting in the increase of $l_{\rm AR}$. This is in agreement with our findings. Their free energy also showed that the bulk energy increases as $\tau$ increases. Therefore, when $\tau$ increases, $l_{\rm AR}$ tends to increase in order to reduce the bulk energy. This is in accord with experimental observations that $l_{\rm AR}$ increases for BCPs with larger $\tau$. Our results, however, show a non-monotonous dependency of $l_{\rm AR}$ on $\tau$. We find that the aspect ratio $l_{\rm AR}$ first increases and then decreases as function of  continuously increasing $\tau$. Such predictions can be tested in future experiments by choosing BCPs that have a larger Flory-Huggins interaction parameter.

\section{Conclusions}

In summary, we systematically studied the mechanism of NP formation from di-block copolymers. We show that a lamellar-forming BCP prefers ellipsoidal NPs with inner lamellae perpendicular to the long axis (called the L$_\perp$ structure), where the solvent preference to one of the two blocks is causing the formation of onion-like NPs with inner lamellae arranged as concentric shells (the C$_\parallel$ structure). When the solvent is neutral towards the two blocks, we showed that more elongated ellipsoidal NPs can be obtained by increasing the A/B interaction parameter $\tau$, or by decreasing the polymer/solvent interaction parameter $\chi$, as well as by increasing the NP size.

On the other hand, when the solvent has a preference towards one of the two blocks, increasing the strength of this preference induces a change in the NP shape from ellipsoidal to spherical, then to onion-like. In a corresponding manner, the inner structure changes from lamellae perpendicular to the long axis to concentric shells. Our results not only are consistent with the current experimental findings, but also predict yet unexplored effects of how the equilibrium shape of BCP-NPs depends on experimental controlled parameters.

In this work, we studied BCP forming nanoparticles for which the BCPs have a lamellar phase in the bulk and the NP shape is mainly ellipsoidal (in 2D systems). BCPs that have a cylindrical structure in the bulk have also been used in the fabrication of nanoparticles. The resulting NPs have an oblate shape and attracted much attention~\cite{Yang16,Ku19,Shin19,Shin20}. However, such theoretical studies have to go beyond a 2D calculation, they lie beyond the scope of our present work and are left to future studies. We hope that the theoretical results presented in this work can be a useful guide for future experiments as well as  applications for nanoparticles form from BCPs.

\bigskip
{\bf Acknowledgement.}~~
This work was supported in part by grants No.~21822302 of the National Natural Science Foundation of China (NSFC), and the NSFC-ISF Research Program, jointly funded by the NSFC under grant No.~21961142020, and the Israel Science Foundation (ISF) under grant No.~3396/19.

\newpage

\clearpage
\vskip 0.5truecm
\centerline{for Table of Contents use only}
\centerline{\bf Formation of Diblock Copolymer Nanoparticles: Theoretical Aspects}
\centerline{\it Yanyan Zhu, Bin Zheng, David Andelman, and Xingkun Man$^*$}

\begin{figure}[h]
{\includegraphics[width=0.5\textwidth,draft=false]{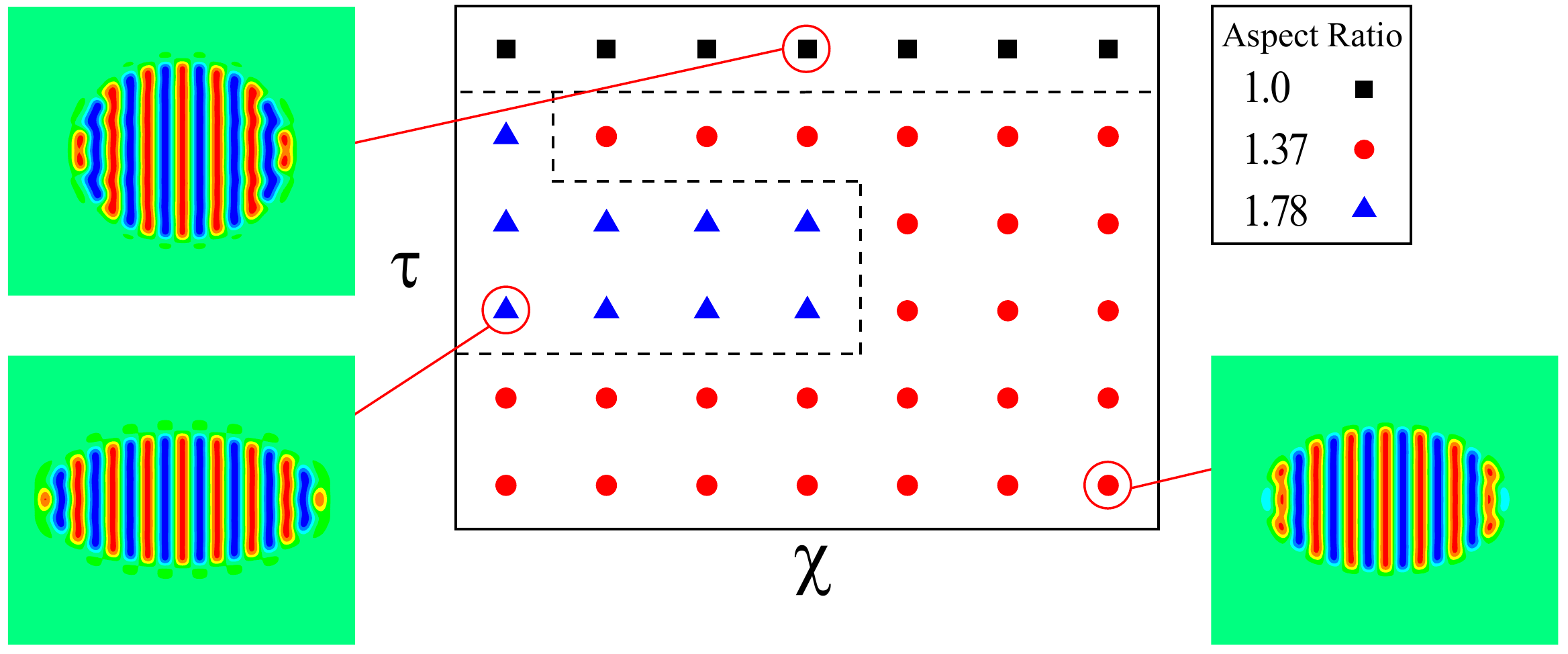}}
\end{figure}

\end{document}